# Molecular dynamics simulations with many-body potentials on multiple GPUs – the implementation, package and performance

Qing HOU[*], Min LI, Yulu ZHOU, Jiechao CUI, Zhenguo CUI, Jun WANG

*Key Lab of Radiation Physics and Technology, Institute of Nuclear Science and Technology, Sichuan University, Chengdu 610064, China*

**\*Corresponding author:** Qing Hou, email: qhou@scu.edu.cn

**Abstract:** Molecular dynamics (MD) is an important research tool extensively applied in materials science. Running MD on a graphics processing unit (GPU) is an attractive new approach for accelerating MD simulations. Currently, GPU implementations of MD usually run in a one-host-process-one-GPU (OHPOG) scheme. This scheme may pose a limitation on the system size that an implementation can handle due to the small device memory relative to the host memory. In this paper, we present a one-host-process-multiple-GPU (OHPMG) implementation of MD with embedded-atom-model or semi-empirical tight-binding many-body potentials. Because more device memory is available in an OHPMG process, the system size that can be handled is increased to a few million or more atoms. In comparison with the CPU implementation, in which Newton's third law is applied to improve the computational efficiency, our OHPMG implementation has achieved a 28.9x~86.0x speedup in double precision, depending on the system size, the cut-off ranges and the number of GPUs. The implementation can also handle a group of small boxes in one run by combining the small boxes into a large box. This approach greatly improves the GPU computing efficiency when a large number of MD simulations for small boxes are needed for statistical purposes.

**Keywords:** molecular dynamics; GPU; CUDA; semi-empirical tight-binding; embedded-atom-model

# 1. Introduction

As one of the most powerful research tools, classical molecular dynamics (MD) has been extensively applied in a wide range of fields in material science. By following the trajectories of all atoms governed by Newton's laws, MD is capable of providing the most detailed evolution information of a dynamic system at an atomistic level as well as providing experimentally observable quantities in combination with statistical physics methods [1]. However, because an MD simulation is computationally quite intensive, especially in cases where the number of atoms involved in a simulation is large [2], it is always desirable to accelerate the performance of MD. One acceleration approach is to develop new algorithms depending on the physics model involved in a given problem. Examples of this approach include the hyperdynamics [3,4] and temperature-accelerated-dynamics (TAD) for the study of atomic diffusion [5] and the method used by Nordlund for the simulation of ion implantation [6]. An alternative acceleration approach is to parallelise the MD simulations, which depends on the development of computer hardware and programming tools [2,7,8]. One such development that has been attracting increasing attention in the past few years because of its effective cost and encouraging performance is the use of a graphic processing unit (GPU) for parallel computing; a number of primary scientific applications demonstrating this development have been summarised by Garland *et al.* [9] and, recently, by Harvey and Fabritiis [10].

Implementations of MD on a GPU have been reported by a number of authors [11-17]. As a primary demonstration of the implementation of MD on a GPU, the GPU-based MD presented by Yang *et al.* was used to calculate the thermal conductivities of solid argon [11]. Because the neighbour lists of atoms were calculated only once and were not updated over time, the implementation of Yang *et al.* is only a very specific application. Implementations of GPU-based MD for a relatively more general application were reported by Anderson *et al.* [14] and van Meel *et al.* [12]. However, only pairwise interactions between atoms were considered in these implementations. It has been recognised that the pairwise potentials are not suitable when the simulation targets involve metal crystals [18,19], which are the systems in which we are interested. It is well known that, for such systems, the many-body nature of the atomic interactions needs to be accounted for to correctly describe the interaction between the atoms. As described in section 2, for many-body potentials, the force on an atom depends not only on the effective electron density on this atom but also on the effective electron densities on its neighbouring atoms, and the densities of these neighbouring atoms are again determined by their neighbouring atoms. Due to this nonlocal feature of the many-body potential, porting MD with many-body potentials onto a GPU is more challenging than with pairwise potentials and provides a good test of the capabilities of the GPU-based parallel computing.

Because a GPU cannot perform independently of a CPU, a GPU computing kernel must be invoked through a host process. The abovementioned implementations of GPU-based MD run on a one-host-process-one-GPU (OHPOG) manner; that is, the GPU kernels in a host process execute on a single GPU. Because the memory available on a GPU is much smaller than that available on a host (for example, the available global memory on a C2050 card is 3 GB, whereas the memory on a host could be a few tens or more GBs), even without considering the computing time, running the GPU kernels on a single GPU limits the system size that an implementation can handle; this is especially true when the particle density of the system is high and an atom has a large number of neighbours, which requires a large memory to store the neighbour lists. An approach to overcome the limitation and also to improve the computing speed is to parallelise the MD simulations on clusters, using MPI for the communication between the host processes [20] with each of the host processes running in an OHPOG manner and processing a subset of the system. The efficiency of using GPUs in this approach could be degraded due to frequent inter-process data exchanges between nodes, especially when a many-body potential is adopted because of the nonlocal feature of the potential. (In Ref. [20], the performance tests were given only for pairwise potentials).

In this paper, we present an implementation of MD simulations with many-body potentials using multiple GPUs in a one-host-process-multiple-GPU (OHPMG) manner, with the original aim of performing MD simulations for systems containing millions of atoms on low-cost hardware platforms, such as workstations, with multiple GPUs. In the OHPMG approach, the abovementioned GPU memory limitation is removed, and the inter-process data exchanges are not needed. (The OHPMG manner can also be implemented on clusters with each node containing multiple GPUs. The inter-node data exchanges in an OHPMG implementation would be less frequent than those in an OHPOG implementation because each node can handle a larger subsystem in the OHPMG approach).

We also note here that a system including a large number of atoms represents two situations that could be encountered in MD simulations. In the first situation, a large simulation box must be used to correctly describe the physics involved in the simulation target; an example of this situation is the simulation of cascade collisions in a material irradiated by energetic atoms [2]. In the second situation, many simulation boxes, each containing only a few hundred or thousand atoms, are required for statistical purposes; an example of this situation is the calculation of the diffusion coefficients of the impurities in a material where assembly averages are required. In this situation, it is not efficient to run the simulations on GPUs in a sequence of the small boxes. In this paper, we address these two situations in the same framework.

**2. Physics model**

An implementation of MD simulations is usually dependent on the physics models. Thus, we describe the physics models involved in our implementation before giving the details of our implementation.

The basic models in MD simulations are the interaction potentials between atoms. It is well known that the atomic trajectories in a MD simulation strongly depend on the interaction potentials. The many-body potentials that are most widely adopted in MD simulations for the structures and dynamics of metals are semi-empirical tight-binding (SETB) potentials [18,21], embedded-atom-model (EAM) potentials [19,22] and their variants. Although these potentials are based on different physical ground, both the SETB and EAM potentials have a similar mathematical form that can be written for an atom $i$:

$$E_i = F(\sum_{j \neq i} \rho^{(ab)}(\mathbf{r}_{ij})) + \frac{1}{2}\sum_{j \neq i} V^{(ab)}(\mathbf{r}_{ij}) \tag{1},$$

where $V^{(ab)}(\mathbf{r}_{ij})$ is the pairwise potential between atoms $i$ and $j$ that are separated by a distance $\mathbf{r}_{ij}$, $F$ is a noncumulative function of the sum $\rho^{(ab)}(\mathbf{r}_{ij})$. In the EAM potential, $\rho^{(ab)}(\mathbf{r}_{ij})$ is understood as the effective electron density at the position of the atom $i$ contributed by the atom $j$, whereas in the SETB potential, $\rho^{(ab)}(\mathbf{r}_{ij})$ is understood as the square of the hopping integral between the atoms $i$ and $j$. The superscripts $a$ and $b$ denote, respectively, the types of the atom $i$ and the atom $j$ for cases in which the system contains different types of atoms. The total cohesive energy of the system is thus

$$E_T = \sum_{i=1}^{N} F(\sum_{j \neq i} \rho^{(ab)}(\mathbf{r}_{ij})) + \frac{1}{2}\sum_{i=1}^{N}\sum_{j \neq i} V^{(ab)}(\mathbf{r}_{ij}) \tag{2},$$

and the force on the atom $i$ is written as

$$\mathbf{f}_i = -\nabla_{\mathbf{r}_i} E_T = -\sum_{j \neq i}(\frac{\partial F(\rho_i)}{\partial \rho_i} + \frac{\partial F(\rho_j)}{\partial \rho_j})\frac{\partial \rho^{(ab)}(r_{ij})}{\partial r_{ij}}\frac{\mathbf{r}_{ij}}{r_{ij}} - \sum_{j \neq i}\frac{dV^{(ab)}(r_{ij})}{dr_{ij}}\frac{\mathbf{r}_{ij}}{r_{ij}} \tag{3},$$

where $\rho_i \equiv \sum_{j \neq i} \rho^{(ab)}(r_{ij})$. The first term in eq. (3) indicates that the force on an atom depends on the effective electron density of this atom as well as the effective electron densities of its neighbouring atoms, which are further determined by these atoms' neighbouring atoms. This feature of the many-body potential makes the parallelisation of the MD computing a complex process compared to the parallelisation of the MD computing with only pairwise potentials.

In practical MD applications, the evolution of the temperature, pressure and volume of the system often must be considered. A number of methods have been developed for controlling the temperature and pressure in MD simulations. Among these methods, the so-called extended system method, in which an additional term is added to the Hamiltonian of the system, is commonly used [23-25]. However, the extended system method could be questionable if the system to be studied is initially far from equilibrium. Thus, for temperature control, we adopt a model proposed by Finnis *et al.* [26] and later applied by Hou *et al.* [27] for simulations of atomic cluster deposition on surfaces. In this model, the atoms exchange energy with an electron gas that has a high thermal conductivity and is thus considered to have a constant temperature. According to this model, a velocity-dependent dumping force is added to the force on each atom, and thus, the movements of the atoms are governed by the Langevin equation

$$m_i \frac{d^2 \mathbf{r}_i(t)}{dt} = \mathbf{f}_i(t) - \frac{1}{\tau_T} \frac{(T_i - T^{(e)})}{T_i} \mathbf{v}_i(t) \qquad (4),$$

where $T_i$ is the instantaneous temperature of atom $i$, $T^{(e)}$ is the electronic temperature, which is assumed to be constant at the given temperature, and $\tau_T$ is the characteristic time describing the rate of the energy exchange between the electrons and atoms.

Regarding pressure control, we adopt the approach proposed by Berendsen *et al.* [28]. In this approach, the coordinates of the atoms and the length of the box are scaled per time step by a factor $\mu$ that is written as

$$\mu = [1 - \frac{\Delta t}{\tau_p}(P_0 - P)]^{1/3} \qquad (5)$$

where $\Delta t$ and $\tau_p$ are the time step and the characteristic time, respectively, $P$ is the instantaneous pressure of the system and $P_0$ is the given external pressure. The instantaneous pressure $P$ is calculated by [1]

$$P = (Nk_B T + \frac{1}{3}\sum_{i=1}^{N} \mathbf{f}_i \cdot \mathbf{r}_i)/V \qquad (6)$$

where $k_B$ is the Boltzmann constant, $T$ is the instantaneous temperature of the system, $V$ is the volume of the box and $N$ the number of atoms.

## 3. Algorithm and implementation of multiple GPU-based MD

Corresponding to the physics models, we have developed a program package MDPSCU for MD simulations. The package was programmed in Fortran 90 and is self-contained and structured, with the routines packed in a number of modules of Fortran 90. The GPU-based codes in MDPSCU were written using CUDA Fortran. The overviews and programming guides for CUDA can be found in the literature [12,14-16,29]. The details about CUDA Fortran can also be found in "CUDA Fortran Programming Guide and Reference" [30]. Here, we focus on algorithms related to GPU computing.

Fig. 1 displays a typical workflow of a GPU-based application using MDPSCU. The workflow is almost the same as that in a CPU-based application. The first phase is to do a serial of initialisations, according to the controlling parameters, on the host and the GPUs.

### 3.1 Initialisation and memory allocation on GPUs

Initializing the devices is to allocate the necessary memory on the devices. Assuming that the number of atoms in the system to be simulated is NA and that the number of GPUs to be used is NG, the system will be partitioned into NG subsystems, and one GPU will handle one subsystem. The process of partitioning the system into subsystems on the GPUs will be described in subsection 3.2. The major arrays to be allocated on the devices include those for the positions, the velocities, the forces, the neighbouring numbers and the neighbour list of the atoms and the electron densities $\rho_i$ of the atoms (denoted respectively by dX, dV, dF, dN, dL and dD in later discussions, where the prefix 'd' represents 'device' and the prefix 'h' represents 'host'). Considering the fact that a system may contain different types of atoms, arrays (denoted by dT) marking the types of the atoms are also allocated. In some applications, the atoms are assigned states. For example, some atoms in a system are not allowed to move. Thus, arrays (denoted by dS) are also allocated to mark the states of the atoms.

The sizes of dV, dF, dN, dT and dS allocated on each GPU are NA(*1+p*)/NG (for convenience, the size of a array means the number of elements of the array; an element of an array for a three-dimensional vector, for instance, the position of an atom, actually contains three components), where *p* is a positive parameter used to adjust the permitted number of atoms that a GPU can handle, considering the fact that the number of atoms in a subsystem actually fluctuates and that a redundancy space is therefore necessary. Because the calculations of the neighbour lists and the forces of a subsystem need the positions and the electron densities of other subsystems, as will be described in subsections 3.2 and 3.3, the sizes of dX and dD allocated on each GPU are NA, the size of the original system. The size of dL, which demands most of the resources, is MN×NA(1+r)/NG, where MN is the maximum permitted number of neighbours of an atom. MN can be adjusted according to the particle density of the system and the cut-off range for the neighbour-list calculations.

Note that the arrays allocated on two GPUs cannot share the same variable names. In practice, the identifiers of the GPUs are used as a suffix of the variables to distinguish the arrays allocated on the different devices. For example, dX1 and dX2 are declared separately for the positions of the atoms handled by the first GPU and the second GPU.

In addition to these major arrays, auxiliary arrays are also allocated when initialising the GPUs. The descriptions of these auxiliary arrays will be given later where they are used.

## 3.2 System partitioning and calculation of neighbour lists

The system is partitioned into subsystems on the basis of the cell structure of the simulation box [1], in which the simulation box is divided into TNC=NC(1)×NC(2)×NC(3) cubic cells with each cell surrounded by 26 neighbouring cells (if a periodic condition is imposed), where the vector NC represents the number of cells in the x, y and z directions. A subsystem is thus constructed by the atoms in a group of cells that are sequentially indexed. For example, the first subsystem contains the atoms in the first TNC/NG cells, the second subsystem contains the atoms in the second TNC/NG cells, etc. Fig. 2 displays the pseudocodes to partition the system. Here and thereafter, lowercase and italic symbols represent temporary variables that were used locally.

The first step is to construct the linked lists (from line H1 to H8). The algorithm for creating the linked lists is normative [1]. Unlike the standard algorithm, the indices of the cells (dIC) that the atoms in are calculated on a GPU. Because the GPU implementation for calculating dIC is quite straightforward, we omit the detailed description for it. In parallel to the calculation of dIC (line G1-G3), some initialisations (line H2) that are needed for constructing the linked lists are performed on the host, although only a very short time is taken by H2. Due to its sequential nature, the generation of the linked lists has to be performed on the host (line H3-H8), with the incidental cost of copying dIC from the device to the host (line G4).

The second step is sorting and aggregating the atoms by the cells (line H9-H22), using the linked lists generated above. The array hGID connects the indices of the atoms after sorting and the indices of the atoms before sorting. For example, the $i$th atom in the sorted atoms is the *hGID(i)*th atom of the original system. The arrays hNAC and hFA, the sizes of which are the number of cells TNC, record the number of atoms contained in the cells and the indices of the first atoms in the cells after sorting, respectively. The arrays hX and hT, which will be used immediately in the next step, store the positions and types, respectively, of the sorted atoms.

The third step is the concurrent execution of calculating the neighbour lists on the GPUs (line G5-G7) and sorting the velocities and other quantities of the atoms on the host (line H22-H28) using the array hGID generated above (In programming, G5-G7 are previous to H22). The calculation of the neighbour lists requires that hX, hT, hFA, hNAC and other box information are copied into the device memory. We use asynchronous copy commands (line G5) for copying these arrays from the host to the devices. Because the GPU kernels (line G6) are originally asynchronous, the host continues to launch copy commands and GPU kernels on the GPUs without waiting for the previously launched copy operations and the GPU kernels to finish; therefore, the host and the GPUs execute concurrently. Whether G6 finishes before H28 or H28 finishes before G6 depends on the system size and the number of GPUs being used.

Fig. 3 displays the pseudocode of the GPU kernel (line G6 in Fig. 2) for the calculation of the neighbour lists for a subsystem. A GPU handles a subsystem, which consists of cells indexed from CELLFROM to CELLTO. The block size for executing the kernel is BLOCKSIZE. A block is devoted to calculate the neighbour lists of the atoms in a cell, with each thread of the block building the neighbour list for one atom. Because BLOCKSIZE is most likely smaller than the number of the atoms in a cell, the atoms in the cells are grouped in NBPC groups, and NBPC blocks are needed to cover all the atoms in the cell, where NBPC=NACmax/BLOCKSIZE and NACmax is the maximum value of hNAC obtained in the second step. The total number of blocks, or the grid size, for executing the kernel on a GPU is thus NBPC*(CELLTO-CELLFROM+1). The block *bid* is then actually devoted to calculate the neighbour lists of the atoms of the *ib*th group in cell *ic*, with *ib*=*bid*-int(*bid*/NBPC)*NBPC and *ic*=*bid*/NBPC+CELLFROM. The index of the atom that the *tid*th thread in the block *bid* addresses is *ia0* in the whole system and is *ia* in the subsystem, with *ia*=*ia0*-dFA(CELLFROM)+1. The neighbours of the atom are searched in cell *ic* and its neighbouring cells. The index *idc* of the neighbouring cells of cell *ic* is most likely not in the interval from CELLFROM to CELLTO, which is why the positions dX and the types dT of the atoms of the whole system are needed on each GPU. In scanning the candidate neighbours, the positions and types of the atoms are copied from the global memories dX and dT into the arrays *spos* and *sityp* that are allocated in the shared memory with size BLOCKSIZE. Again, because BLOCKSIZE is most likely smaller than the number dNAC(*idc*) of the atoms in cell *idc*, a neighbouring cell is scanned starting from the atom *jafrom* for *ns* times, with the index *jafrom* shifting up by BLOCKSIZE for each time. The number of the neighbour atoms is accumulated in *n* and finally assigned to the global memory dN. Because *jafrom* is the atom index in the whole system, the indices of the neighbour atoms recorded in dL for the *ia*th atom in the subsystem are the atom indices in the whole system.

Independent of whether the G6 has finished, the host launches the asynchronous copy command G7, which copies the segments of hV, hS, etc. into the device memories. For example, the array segment hV(1:hIAD(1)) is copied into the first GPU, where hIAD(1) is the index of the last atom in the last cell handled by this GPU. After launching the copy commands, the host continues other operations without waiting for the finish of the copy operations.

## 3.3 Predictor-corrector scheme and force calculations

With the partitioned system, the stepwise integrations of the dynamics equations for the subsystems are performed on GPUs. A predictor-corrector scheme is used for the stepwise integration [1]. The pseudocode is displayed in Fig. 4. Algorithmically, the GPU implementation (line G8-G14) for predicting the positions and velocities is straightforward. Each thread predicts the position and the velocity of one atom. Note that, corresponding to the thread *bid* in the block *bid* on a GPU, the atom index for the velocity prediction is *ia=bid*\*BLOCKSIZE+*tid*, the index of the atom in the subsystem, whereas the index for the position prediction is *ia0* =*ia*+dFA(CELLFROM), the index of the atom in the whole system.

After the predictions of the positions and velocities, the positions are combined by asynchronously copying segments of dXs from the GPUs into the host array hX and then copying hX back into dXs (line G15-G17) because the dD and dF for an atom to be calculated are most likely contributed from the neighbouring atoms in different subsystems.

According to eq. (3), the force on an atom depends not only on the effective electron density of itself but also on the effective electron densities of its neighbouring atoms, whereas the electron densities of the neighbouring atoms are in turn contributed by their neighbours. Due to the nonlocal feature of the force calculation, the force calculations require two GPU kernels, G18 and G22. Fig. 5 displays the pseudocode of the GPU kernel G18, which calculates the effective electron densities of the atoms. Because the total number of threads for the kernel execution BLOCKSIZE\*GRIDSIZE is probably smaller than the number of the atoms in a subsystem (NAPD), a thread is likely devoted to the calculation of the electron densities of *nl*=NAPD/(BLOCKSIZE\*GRIDSIZE)+1 atoms, with the indices of these atoms separated by BLOCKSIZE\*GRIDSIZE. After the atom index *ia* in the subsystem and the index *ia0* in the whole system is determined, the neighbour list of this atom is scanned to calculate the contributions of the neighbours to the electron density dD(*ia0*). The contribution from the atom *ja0* to the atom *ia0* is obtained by interpolation from $\rho_{ity,jty}(r)$, a lookup table of the electron density contributed from an atom of the type *jty* to an atom of the type *ity*. The lookup table $\rho_{ity,jty}(r)$ is created once on a CPU and copied to the device upon starting the MD simulation. For the same reason used for G15-G17, the electron densities dD obtained on each GPU need to be combined (line G18-G21) for the following force calculations.

Fig. 6 displays the pseudocode of the GPU kernel G22 for the force calculations. The workflow of G22 is similar to that of G15. The force contributions from the many-body potential (the first term in eq. (3)) and the pairwise potential, respectively indicated by $f_1$ and $f_0$, are calculated; $\partial\rho(r)/\partial r$ and $\partial V(r)/\partial r$ are obtained by interpolations from corresponding lookup tables, which are also created once on a CPU and copied to the devices upon starting the MD simulation. In addition, a shared memory *ps* is assigned for each block to store the contributions to the virial pressure from the forces calculated in a thread. The virial partial pressure contributed from the forces calculated in a block, dP, is then obtained by accumulating *ps*. Because the mechanism for the communication between blocks is missing, the total pressure will be obtained by copying dP from the devices to the host and summing all the partial pressures on the host.

The calculations of the friction forces appearing in the Langevin eq. (4) are also implemented on the GPU. Because the GPU implementations to correct velocity and calculate the friction forces appearing in the Langevin eq. (4) are straightforward, we omit these implementation details.

## 4. Performance evaluation

The performance evaluation of our GPU implementations was performed for the neighbour-list calculations and the force calculations as well as for the full MD simulations separately. A number of hardware and software factors may influence the performance of the implementations. The benchmark platform used for the evaluation was a server running on the operation system Windows Server-2008 R2. The server contains two 2.8 GHz Intel Xeon X5660 processors and 32 GB of RAM. Six NVIDIA TESLA C2050 cards, with 3 GB of global memory, were plugged in specifically for the GPU computing. The performance may be dependent of the versions of the CUDA driver and the Fortran compiler. The version of the CUDA driver was 8-17-12-9573, and the compiler was PGI Visual Fortran version 12.5. The optimisation option in compiling the codes was set as "Maximise Speed", which affected only the host codes.

The potential actually used in the evaluation was a SETB potential, in which the embedment function $F$ is simply $-\sqrt{\sum \rho^{(ab)}(\mathbf{r}_{ij})}$, given by Ackland and Thetford for the metal W [31]. Because the computing times of both the CPU implementations and the GPU implementations strongly depend on the number of candidate neighbours in the neighbour-list calculations and on the number of neighbours in the force calculations, the evaluation was conducted using different cut-off distances (RCUT in Fig. 3 and Fig. 5-6). Additionally, the performance of the GPU kernels also depends on the combinations of the system size, the block size and the grid size for executing the kernels. The system size from 2000 to 2 million atoms was chosen. In practice, it is not meaningful to adjust the block size case-by-case for optimisation performance. Thus, we fixed the block size to be 128 in the neighbour-list calculations and 256 in the force calculations.

In all the benchmark calculations, with the exception of the single-precision variables *pos* and *spos* in the kernel G6 (Fig. 3) for the neighbour-list calculations, all floating variables were in double precision.

**4.1 Neighbour-list calculations**

Fig. 7 displays the runtimes of the GPU-based and CPU-based calculations of the neighbour-lists vs. the system size (measured by the number of the atoms NA). The computing time of the GPU-based calculations is actually the time required to finish all the host and GPU steps in Fig. 2. The order in which step H29 and step G7 finish depends on the system size NA and the number of GPUs. Note that the sorting of the velocities and other quantities of the atoms, which is performed by steps H23-H28 in the GPU-based implementation, is not needed in the CPU implementation. The cut-off range RCUT was chosen as (a) 2.76 and (b) 3.22 in the unit of lattice length, respectively. For different system sizes, because the box size is not exactly an integer multiple of the cut-off range, the average number $NAC_{av}$ of atoms in a linked cell varies from 144 to 157 for RCUT=2.76; therefore, the number of the candidate neighbours to be searched for an atom varies from 3888 to 4239. For RCUT=3.22, $NAC_{av}$ =250, and the number of the candidate neighbours is 6750 for all system sizes.

If only neighbour lists are calculated, the largest subsystem size among the in-tests with which one GPU (TESLA C2050) can deal is 686000 atoms. As will be shown later, for full MD simulations, the maximum number of atoms that one GPU can handle is smaller than this value because other memory expenses are necessary. For NAs larger than 50000, the computing times are approximately linearly correlated with the system size.

Fig. 8 displays the speedup of the GPU-based calculations vs. the CPU-based calculations. For the smallest system size NA=2000 of the in-tests, the speedup is 5.7x for using one GPU in both the cases of RCUT=2.76 and RCUT=3.22. Using more than one GPU is not further beneficial. For NA=16000, the speedup is greatly increased to 65x for RCUT=2.76 and 67x for RCUT=3.22. However, there is still no benefit to use multiple GPUs. For NA from 56000 to 686000, a speedup of 79x~94x is achieved in the case where RCUT=2.76 and 92x~110x in the case where RCUT=3.22 by using one GPU (NG=1).

Using multiple GPUs begins to be beneficial at system sizes NA>56000. In the case of RCUT=2.76, the speedup is 119x ~137x for NG=2 (with the maximum system size NA=1024000 for NG=2), 142x~245x for NG=4 and 142x~286 for NG=6, respectively. In the case of RCUT=3.22, in which the number of the candidate neighbours increases, the speedup is 150x~195x for NG=2, 202x~334x for NG=4 and 225x~396 for NG=6. From Fig. 2, it is obvious that the gain of using multiple GPUs against using a single GPU depends on the quotient of the runtime of the steps H1-H22 and on whether the time taken by the steps G5-G7 can hide the time taken by the steps H23-H29, or vice versa. With an increasing system size and neighbour searching range, the subsystem size handled by each GPU correspondingly increases, the computing time taken by the steps G5-G7 increases, and there is therefore a gain of increasing the number of GPUs, for example, in the cases of RCUT=3.22, NA>54000 and RCUT=2.76, NA>686000.

## 4.2 Force calculations

Fig. 9(a) displays the runtimes of the force calculations. The runtime of the GPU-based calculation is the average of the runtimes of the steps G18-G22 in Fig. 4 over 100 runs, whereas the runtime of the CPU-based calculations is the average runtime over 4 runs. The cut-off range for neighbour searching was chosen as 2.76 times the lattice length, and the cut-off range for the force calculations was 2.3 times the lattice length, respectively. The runtime is approximately linear related to the system size for both the CPU-based calculations and the GPU-based calculations using one GPU. For the force calculations using multiple GPUs, good linear relationships of the runtime vs. the system size are observed only for the system sizes of the in-tests larger than 16000. Fig. 9(b) shows the speedup of the GPU-based force calculations vs. the CPU-based force calculations. For the smallest system size NA=2000 of the in-tests, a speedup of 15.8x was reached for using one GPU, whereas the speedup is 7.92x both for using two and four GPUs. It is likely that the overhead time for launching the kernels (G18 and G22) on the GPUs and the copy operations (G19-G21) occupy a relatively large portion of the whole runtime for small systems and that using multiple GPUs would increase the overhead time. For the system size

NA>16000, the advantage of using multiple GPUs becomes apparent. The speedups of 17.4x~23.2x, 34.0x~42.7x and 56.7x~89.2x were reached for the numbers of GPUs NG=1, 2 and 4, respectively. For NG=6, the speedup at NA=16000 is the same 56.7x as for NG=4. As the system size increases further, the speedup is increased to 88.4x~132.6x. Fig. 9(c) shows the relative speed of using multiple GPUs over using one GPU (note that the largest system size handled by using one GPU is 686000). Clearly, by increasing the system size, the scaling property tends to be linear with the number of the GPUs.

In some MD applications, the pressure may not be of concern; therefore, the steps G22-19, G22-24 and G22-25 in Fig. 6 could be excluded, and a further speedup could be gained. The tests performed with the same calculation parameters above but without the inclusion of the pressure calculation show that speedups of 18.2x~34.2x, 52.6x~70.7x, 107.9x~125.6x and 152.1x~200x were reached for NG=1, 2, 4 and 6, respectively, for NA >128000.

**4.3 Full MD simulations**

Fig. 10 displays the runtimes and speedup of the GPU-based MD simulations vs. the CPU-based MD simulations. In these MD simulations, the update of the neighbour list was performed for every ten time steps. Note that, unlike in the above independent evaluations of neighbour-list and force calculations, the application of Newton's third was added in the CPU implementation and the runtime was reduced by approximately a half in the CPU-based MD simulations. The calculation of the pressure was not included in the GPU-based MD simulations. Again, with an increasing system size, the portion of the overhead time plus the time taken by the sequential parts in the process decreases. The reached speedups are 12.7x~18.2x for NA>1600 and NG=1, 28.9x~36.2 for NA>56000 and NG=2, 47.8x~65.9x for NA>56000 and NG=4, and 68.3x~86.0x for NA>128000 and NG=6. The relative speed of using multiple GPUs over using one GPU for NA from 2000 to 686000 is shown in Fig. 10(c). Furthermore, with an increasing system size, the relative speed increases and tends to be stable at a value. For example, the average speedup of using two GPUs over using one GPU is approximately 1.88 for NA from 54000 to 686000. The maximum speed ratio of using multiple GPUs over using one GPU is 3.5 and 4.6 for NG=4 and NG=6, respectively, at NA=686000.

**4.4 MD simulations of many small boxes**

It has been shown above that the gain of GPU-based MD simulations is not significant if the system size is smaller than ten thousand, especially when multiple GPUs are used, due to the relatively increasing portion of the overhead times and the time taken by the sequential parts in the process. However, there are MD applications, in which many simulation boxes are required for statistical purposes, for example, the calculations of the reflection coefficients of low-energy atoms on surfaces [32]. Many small simulation boxes can be combined into a large box but should be kept independent from each other. The multiple-box-in-one-run scheme is realised in MDPSCU at the partitioning stage described in section 3.2 by shifting the indices of the cells in the *I*th box by *I*xTNC and by requiring that only the cells in the same box can be the neighbouring cells. A demonstrating case, in which one hundred simulation boxes each contain two thousand atoms, was considered. Table 1 compares the runtimes of the CPU and GPU calculations. The GPU calculations were performed for one-box-in-one-run and one-hundred-boxes-in-one-run schemes, respectively. In the case of the one-box-in-one-run scheme, the speedup relative to the CPU calculation is only 5.9, and the speedup tended to be less if multiple GPUs were used. However, the efficiency of the GPU calculations, either using a single GPU or multiple GPUs, is greatly improved in the one-hundred-boxes-in-one-run scheme.

**5. Concluding remarks**

We have described a one-host-process-multiple-GPU (OHPMG) implementation for MD simulations. In our experience, the more general an implementation is, the less computationally efficient this implementation will usually be. Thus, our implementation is aimed at MD with many-body potentials of the types of EAM or semi-empirical tight-binding potentials, which are widely used to simulate dynamics of metals. Compared to the CPU implementation on our benchmark platform, the maximum speedup of the in-tests achieved by the OHPMG implementation is 86x for full MD simulations in double precision. For MD simulations concerning a large number of small boxes, a multiple-box-in-one-run scheme was presented, greatly improving the performance of the GPU calculations for small boxes. The results are encouraging. In our OHPMG implementation, the data exchanges between the GPUs are conducted through the operations of copying data from the device to the host, combining the data on the host and then copying the data from the host to the device. After CUDA 4.1, the GPUs can directly exchange data. However, we did not realise the direct data exchange between the GPUs, most likely due to the compatibility between our platform and the compiling tools. It is anticipated that, with developments in hardware and software, further improvements can be achieved.

Achieving a large speedup is not the only advantage of the OHPMG implementation. Because the available memory on a GPU is small compared to the available host memory, the system size that can be handled by using one GPU is limited. In the OHPMG system, this limitation is overcome. With the increased available device memories, MD simulations containing a few million or more (depending on, for example, the cut-off ranges and density of atoms) atoms can be conducted on a workstation. In principle, the OHPMG scheme can also be implemented on clusters using MPI with each node containing multiple GPUs. Because each node can handle a large subsystem by OHPMG, for a given system size, the data exchanges between the nodes would be less frequent in OHPMG than those when only one GPU is used on one node. Thus, gains in the computational efficiency are expected for implementing OHPMG on clusters.

A description for the details and usage of the program package MDSCU is in preparation and will be submitted for publication in the future.

**ACKNOWLEDGMENTS**: This work was partly supported by National Magnetic Confinement Fusion Program of China (2013GB109000) and the National Natural Science Foundation of China (Contract No. 91126001 and 11175124). We thank professor Marc Hou of Universite Libre de Bruxelles for him providing the primary

MD codes from which we developed the package MDSCU.

**Figure Captions**

**Fig. 1.** Typical workflow of a GPU-based MD application. The boxes with a solid border denote the CPU-only parts. The dashed borders denote the parts related to the data exchange between the host and GPUs. The grey-coloured boxes denote the parts related to GPU computing.

**Fig. 2.** Pseudocodes and workflow for the system partitioning and the neighbour-list calculations, corresponding to the mainstream step A2 and step A5 in Fig. 1. The boxes with a solid border denote the CPU codes. The dashed borders denote the codes of the data exchange between the host and GPUs. The grey-coloured boxes denote the GPU kernels. The italic lowercase symbols denote the variables used locally. This convention is also applied to the pseudocodes displayed later.

**Fig. 3.** A detailed pseudocode of the GPU kernel (step G6 in Fig. 2) for the neighbour-list calculations.

**Fig. 4.** Pseudocodes and workflow for the force calculations as well as the predicting and correcting scheme, corresponding to the steps from A6 to A8 in the mainstream shown in Fig. 1.

**Fig. 5.** A detailed pseudocode of GPU kernel (step G18 in Fig. 4) for the calculations of the electron densities of atoms.

**Fig. 6.** A detailed pseudocode of GPU kernel (step G22 in Fig. 4) for the calculations of the forces on atoms.

**Fig. 7.** Comparison of the runtimes of the CPU implementation and the GPU implementation for the system partitioning and the neighbour-list calculations, corresponding to the workflow in Fig. 2. The cut-off ranges for the neighbour-list calculations are (a) 2.76 times the lattice length and (b) 3.33 times the lattice length.

**Fig. 8.** Speedup of the GPU implementation vs. the CPU implementation for the system partitioning and the neighbour-list calculations. The cut-off ranges for the neighbour-list calculations are (a) 2.76 times the lattice length and (b) 3.33 times the lattice length.

**Fig. 9.** Comparison of performance of the GPU implementation vs. the CPU implementation for the force calculations, corresponding to the steps G18-G22 in Fig. 4. The cut-off range for neighbour-list calculations is 2.76 times the lattice length, and the cut-off range for the force calculations is 2.3 the lattice length. (a) The runtime; (b) the speedup; (c) the relative speed of using multiple GPUs over using a single GPU.

**Fig. 10.** Comparison of performance of the GPU implementation vs. the CPU implementation for full MD simulations, corresponding to the steps A3-A11 in Fig. 1. The cut-off ranges for the neighbour-list calculations and the force calculations are 2.76 and 2.3 times the lattice length, respectively. (a) The runtime; (b) the speedup; (c) the relative speed of using multiple GPUs over using a single GPU.

**Table 1.** The runtime per box per time step and speedup of performing MD simulations for a large number of small boxes in one run.

**Figure 1**

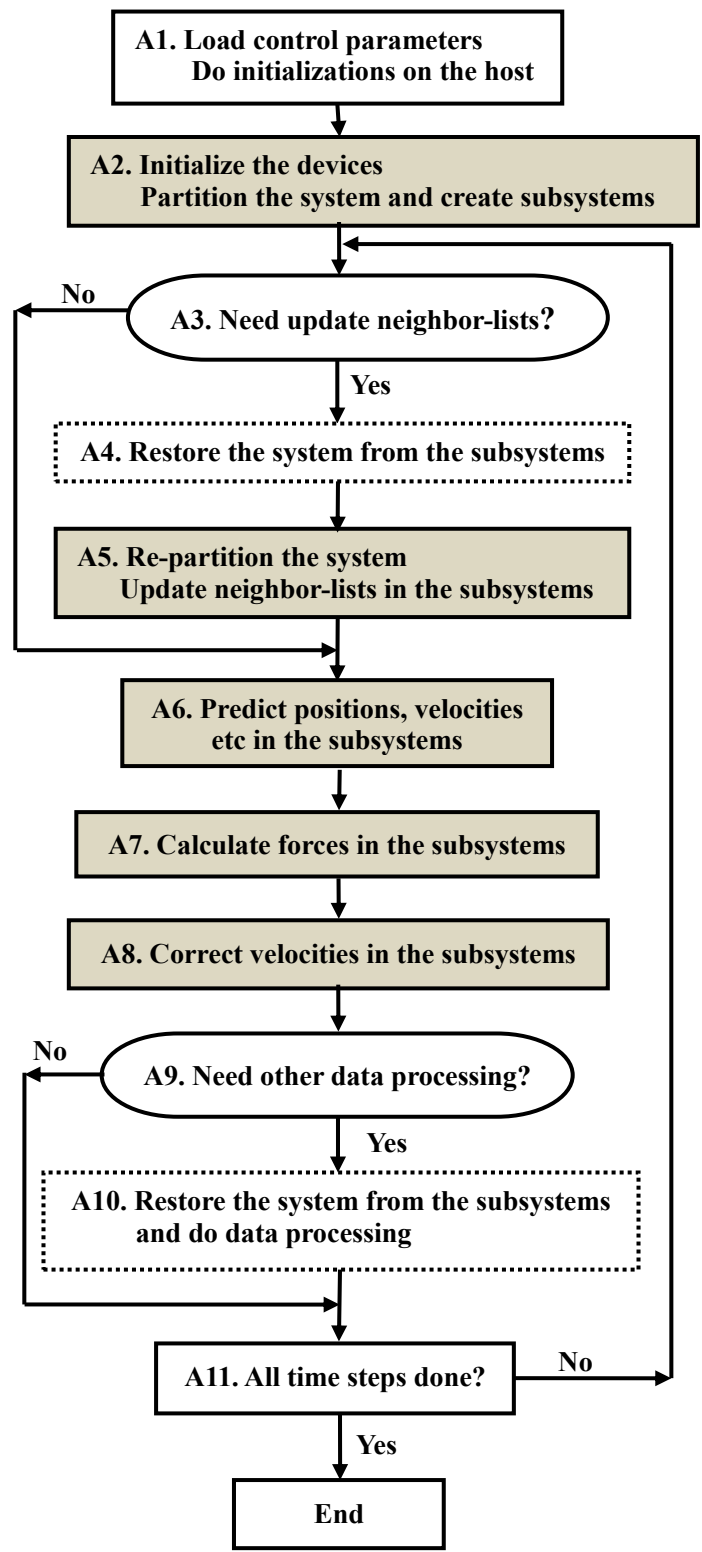

**Figure 2**

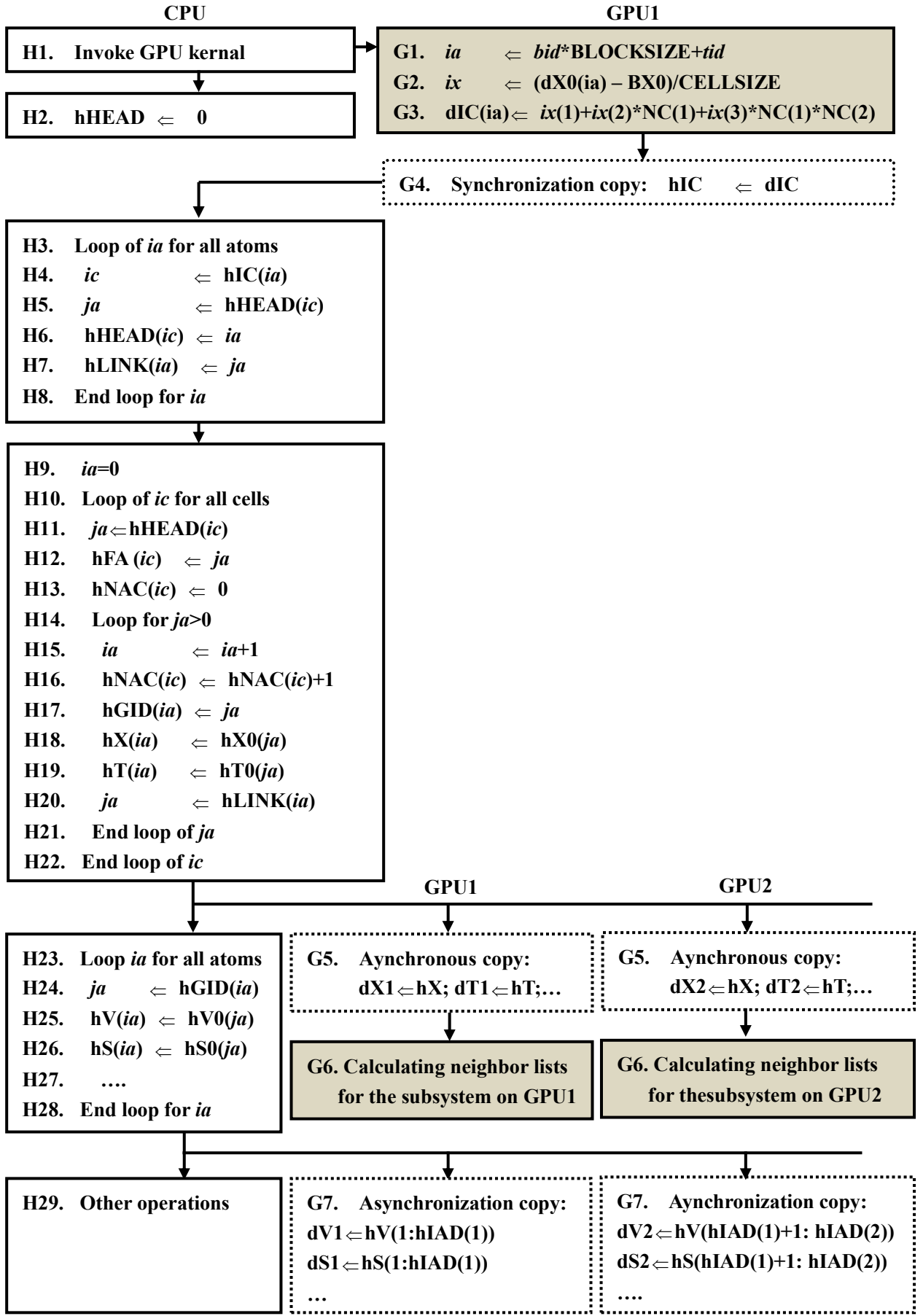

**Figure 3**

```
G6-1.   ic     ⇐ bid/NBPC + CELLFROM
G6-2.   if ic > CELLTO then return
G6-3.   ib     ⇐ bid - int(bid/NBPC)*NBPC
G6-4.   ia0    ⇐ tid + dFA(ic) + ib*BLOCKSIZE
G6-5.   ia     ⇐ ia0 – dFA(CELLFROM)+1
G6-6.   pos    ⇐ dX(ia0)
G6-7.   ity    ⇐ dT(ia0)
G6-8.   n      ⇐ 0
G6-9.   Loop for i from 1 to 27
G6-10.      idc  ⇐ get the id of the ith neighboring cell of cell ic
G6-11.      ns   ⇐ dNAC(idc)/BLOCKSIZE
G6-12.      Loop for j from 1 to ns
G6-13.         jafrom ⇐ dFA(idc) + (j-1)* BLOCKSIZE
G6-14.         jato   ⇐ iafrom + BLOCKSIZE
G6-15.         spos(tid) ⇐ dX(tid+iafrom)
G6-16.         sity(tid) ⇐ dT(tid+iafrom)
G6-17.         Synchronize threads
G6-18.         Loop for ja0 from jafrom to jato
G6-19.            k ⇐ ja0 - jafrom + 1
G6-20.            r ⇐ minimum image of |pos-spos(k)|
G6-21.            if r < RCUT(ity, sity(k)) then
G6-22.               n ⇐ n + 1
G6-23.               dL(ia, n) ⇐ ja0
G6-24.            end if
G6-25.         End loop for ja0
G6-26.      End loop for j
G6-27.   End loop for i
G6-28.   dN(ia) ⇐ n
```

**Figure 4**

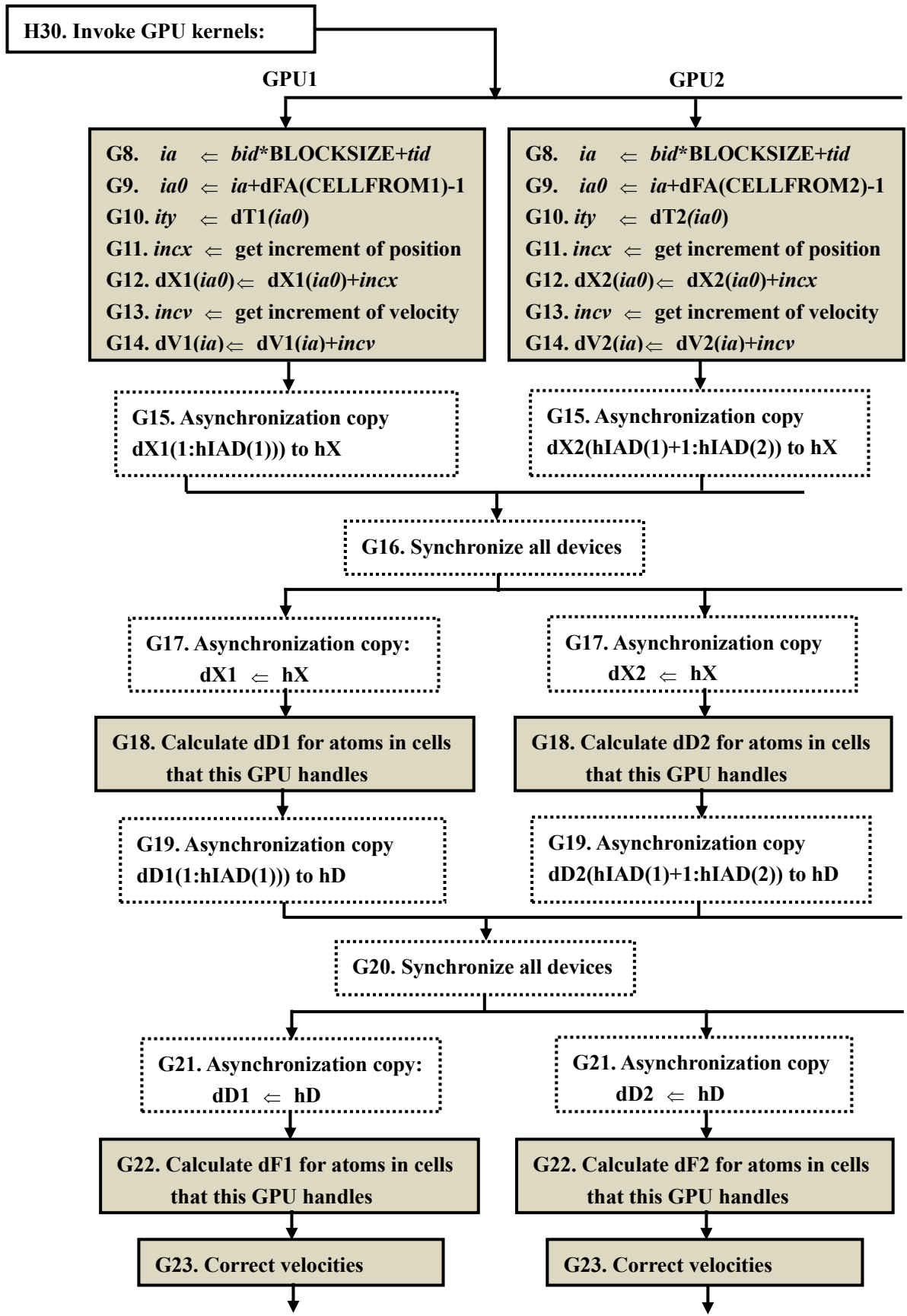

**Figure 5**

```
G18-1.   nl ⇐ NAPD/( BLOCKSIZE*GRIDSIZE)+1
G18-2.   Loop for i from 1 to nl
G18-3.      shift ⇐ (i-1)*( BLOCKSIZE*GRIDSIZE)
G18-4.      ia ⇐ (bid-1)* BLOCKSIZE+bid + shift
G18-5.      ia0 ⇐ ia + dFA(CELLFROM)
G18-6.      pos ⇐ dX(ia0)
G18-7.      ity ⇐ dT(ia0)
G18-8.      Loop for j from 1 to dN(ia)
G18-9.         ja0 ⇐ dL(j)
G18-10.        jty ⇐ dT(ja0)
G18-11.        r ⇐ minimum image of |pos-dX(ja0)|
G18-12.        if r < RCUT(ity, jty(ja0)) then
G18-13.           calculate density d by interpolating of $\rho_{ity,jty}(r)$ table
G18-14.           den ⇐ den + d
G18-15.        end if
G18-16.     End loop for j
G18-17.     dD(ia0) ⇐ den
G18-18.  End loop for i
```

**Figure 6**

```
G22-1.   nl ⇐ NAPD/( BLOCKSIZE*GRIDSIZE)+1
G22-2.   sp(tid) ⇐ 0
G22-3.   Loop for i from 1 to nl
G22-4.      shift ⇐ (i-1)* /( BLOCKSIZE*GRIDSIZE)
G22-5.      ia ⇐ (bid-1)* BLOCKSIZE+tid + shift
G22-6.      ia0 ⇐ ia + dFA(CELLFROM)
G22-7.      pos ⇐ dX(ia0)
G22-8.      den ⇐ dD(ia0)
G22-9.      ity ⇐ dT(ia0)
G22-10.     Loop for j from 1 to dN(ia)
G22-11.        Ja0 ⇐ dL(j)
G22-12.        jty ⇐ dT(ja)
G22-13.        r ⇐ minimum image of |pos-dX(ja0)|
G22-14.        if r < RCUT(ity, jty(ja0)) then
G22-15.           denj ⇐ dD(ja0)
G22-16.           calculate pairwise force f0 by interpolating
G22-17.           calculate by many-body force f1 by den and denj
G22-18.           f ⇐ f + f0 + f1
G22-19.           sp(tid) ⇐ sp(tid) + f*pos
G22-20.        end if
G22-21.     End loop for j
G22-22.     dF(ia) ⇐ f
G22-23.  End loop for i
G22-24.  Synchronize threads
G22-25.  dP(bid) ⇐ accumulating sp for all threads
```

**Figure 7a**

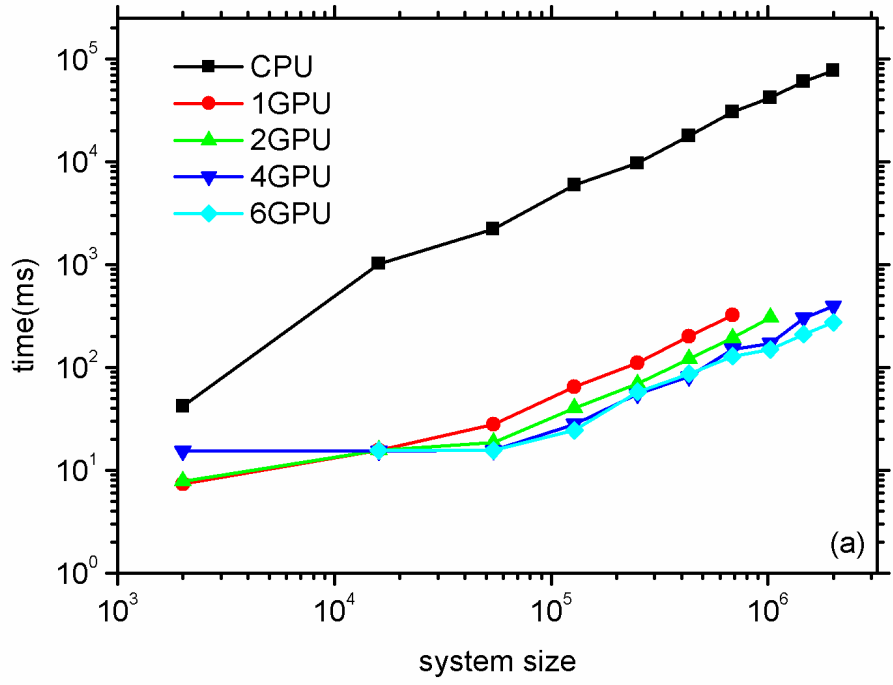

**Figure 7b**

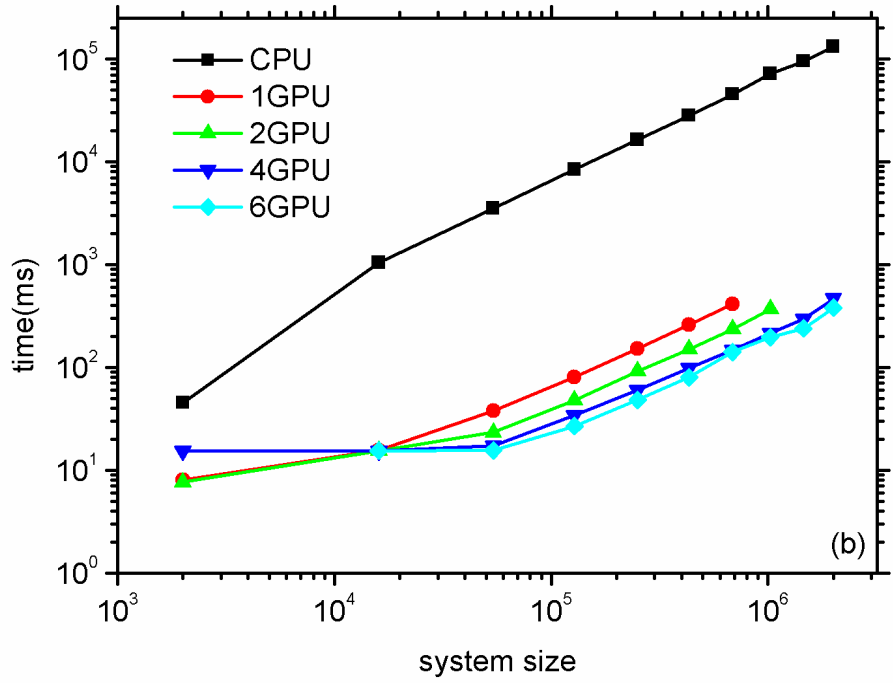

**Figure 8a**

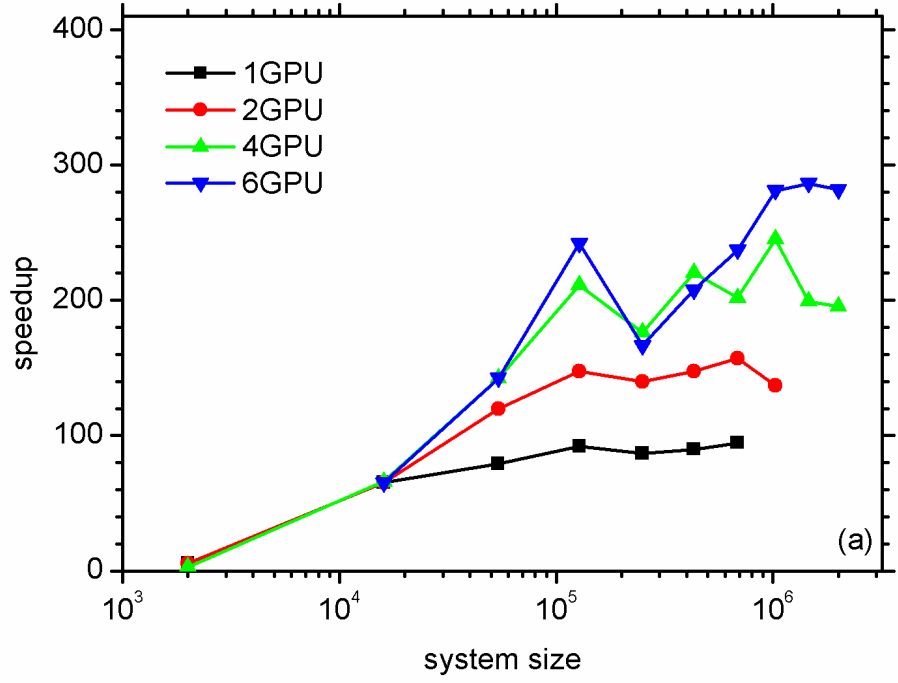

**Figure 8b**

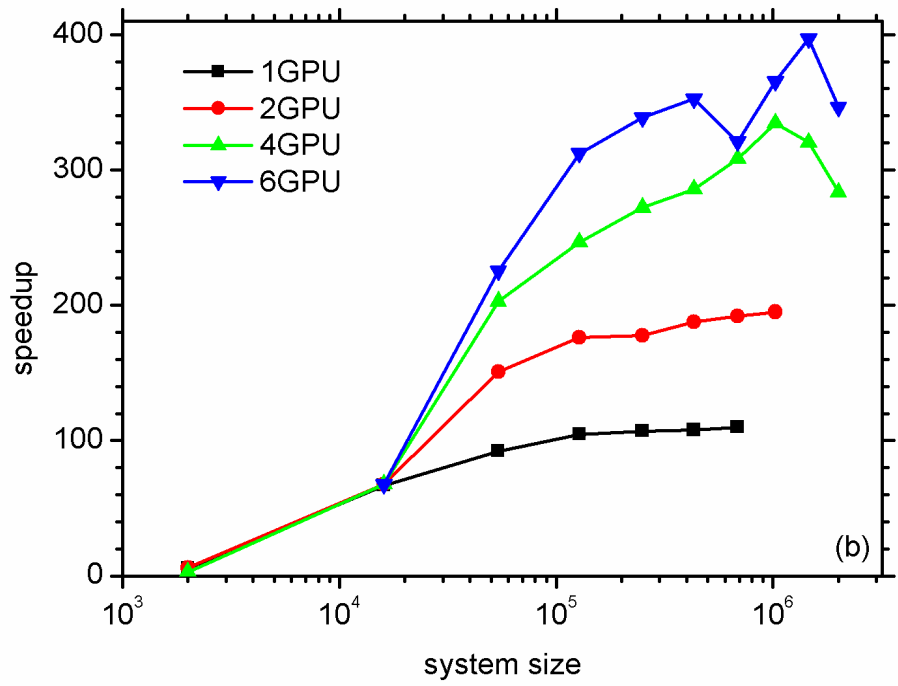

**Figure 9a**

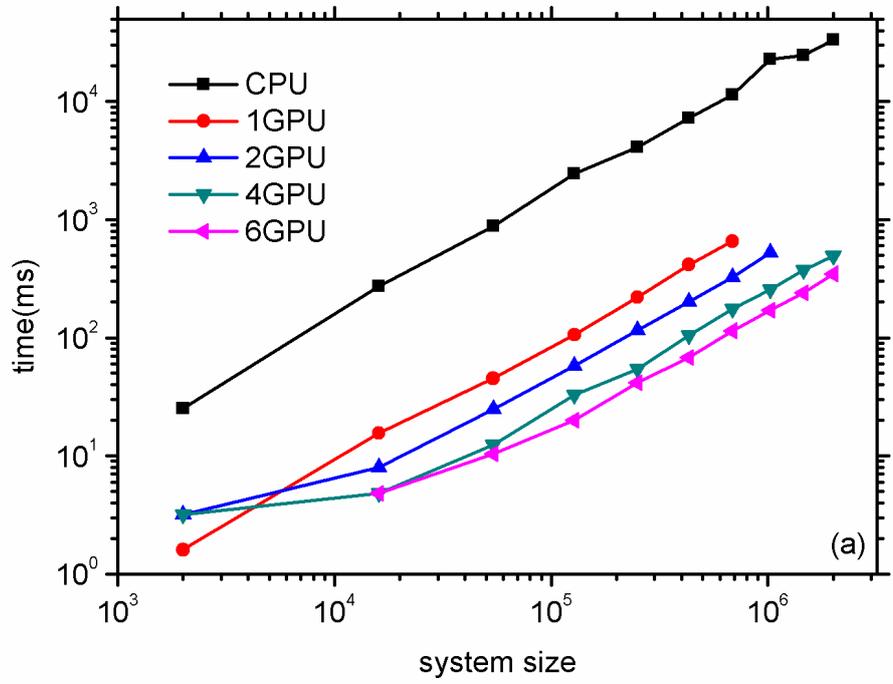

**Figure 9b**

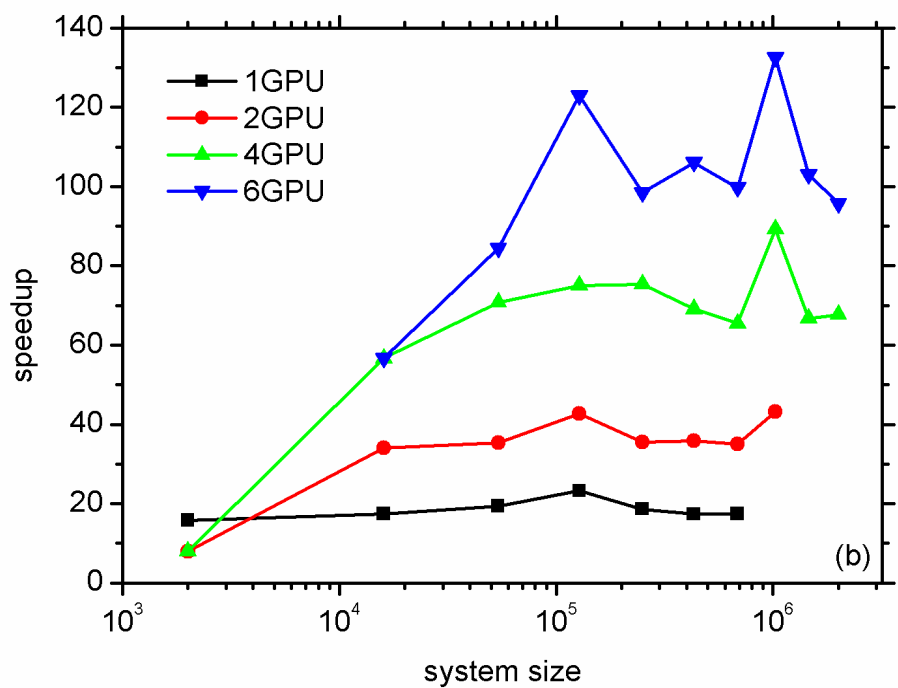

**Figure 9c**

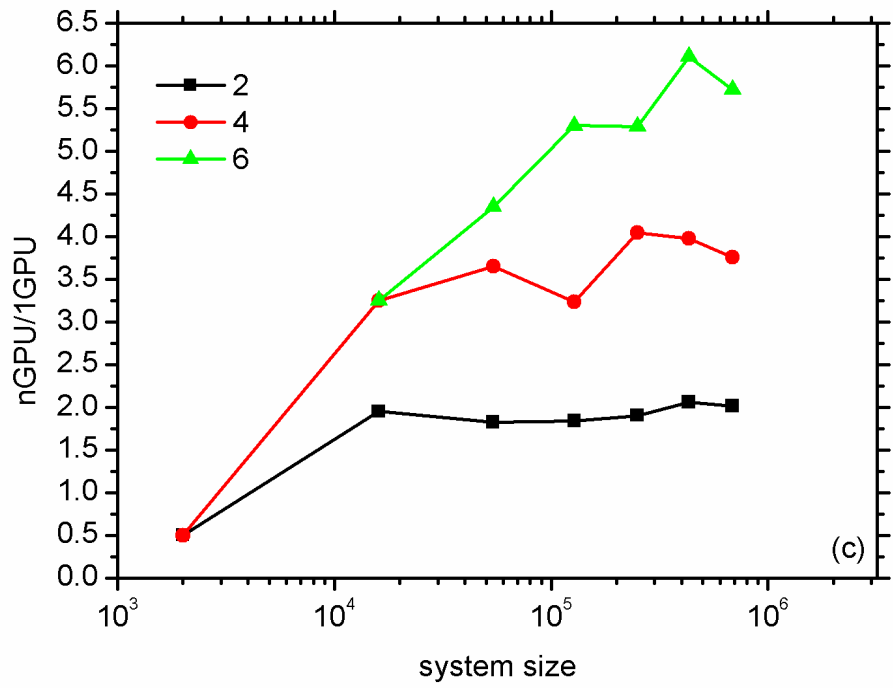

**Figure 10a**

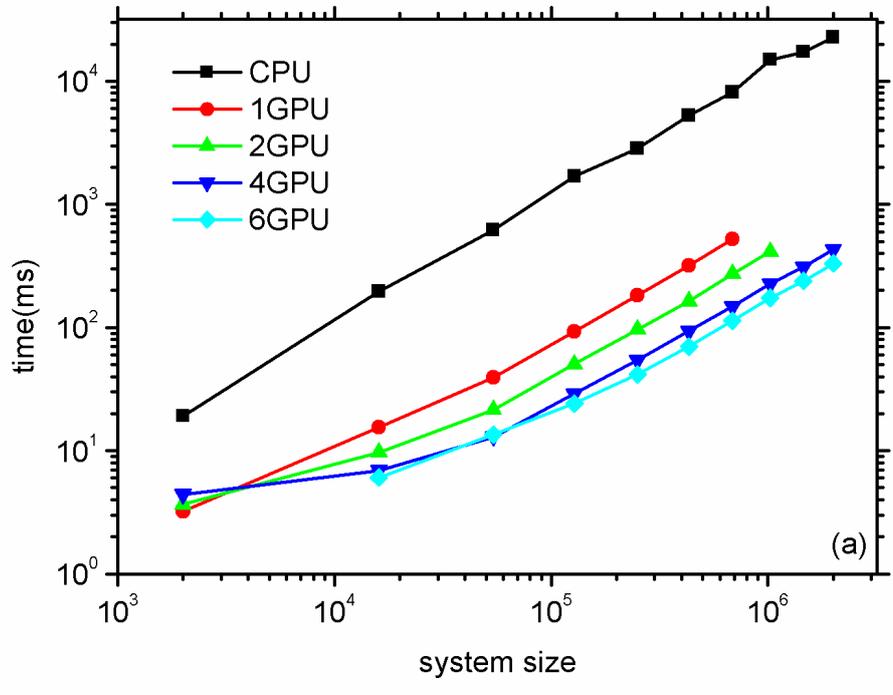

**Figure 10b**

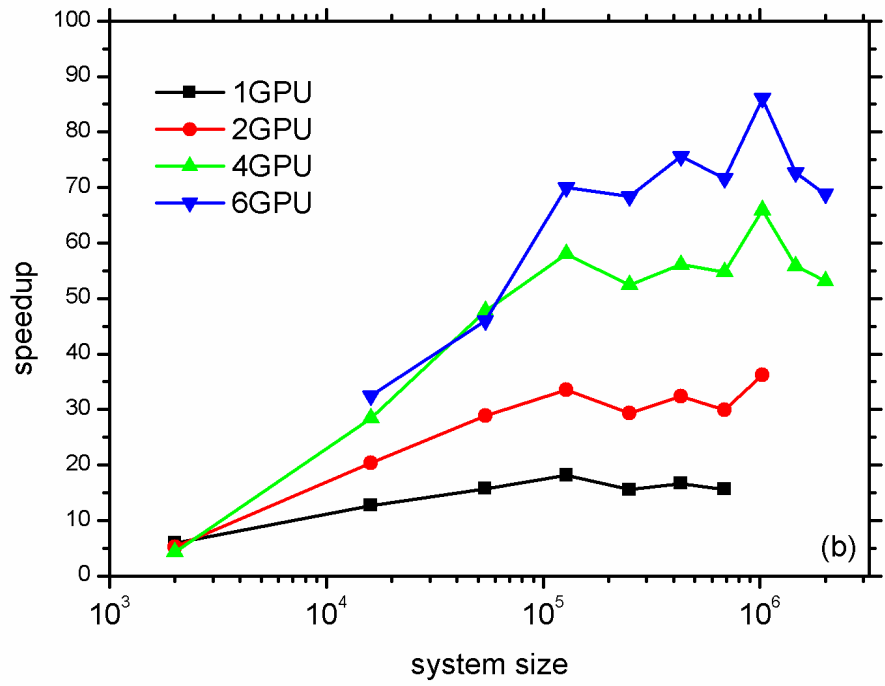

**Figure 10c**

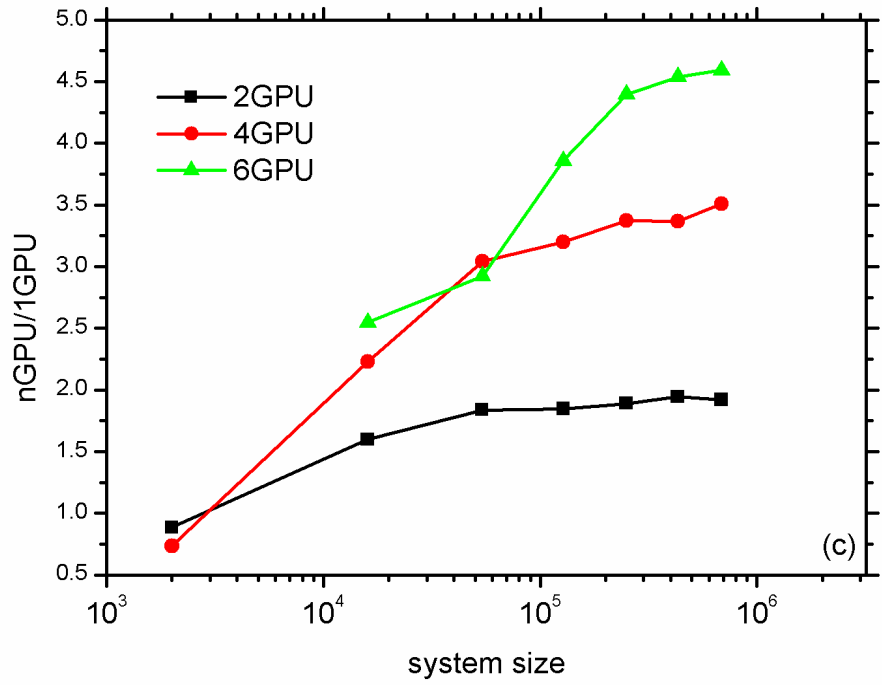

**Table 1.** The runtime per box per time step and speedup of performing MD simulations for a large number of small boxes in one run.

|  | CPU | 1 box one run | | | 100 boxes one run | | | |
| --- | --- | --- | --- | --- | --- | --- | --- | --- |
|  |  | 1GPU | 2GPU | 4GPU | 1GPU | 2GPU | 4GPU | 6GPU |
| time(ms) | 19.88 | 3.24 | 3.66 | 4.42 | 1.34 | 0.70 | 0.40 | 0.30 |
| speedup |  | 6.14 | 5.43 | 4.50 | 14.85 | 28.59 | 49.71 | 65.4 |